\documentstyle[11pt,paspconf,epsf]{article}

\def\sles{\lower2pt\hbox{$\buildrel {\scriptstyle <}
   \over {\scriptstyle\sim}$}}
\def\sgreat{\lower2pt\hbox{$\buildrel {\scriptstyle >}
   \over {\scriptstyle\sim}$}}
\def\undertext#1{$\underline{\smash{\hbox{#1}}}$}

\def\sgreat{\lower2pt\hbox{$\buildrel {\scriptstyle >}
   \over {\scriptstyle\sim}$}}
\def\mj{$\,{\rm M}_{\rm J}\,$}
\def\mjj{{\rm M}_{\rm J}\,}

\def\etal{{\it et~al.}}
\def\mo{$M_\odot\,$}
\def\lbol{L$_{bol}\,$}

\def\Dwa{$\,$\uppercase\expandafter{\romannumeral5}$\,$}
\def\rj{R$_{\rm J}\ $}

\begin{document}

\title{Extrasolar Giant Planet and Brown Dwarf Models}

\author{A. Burrows}\affil{Department of Astronomy, University of Arizona,
        Tucson, AZ 85721}
\author{W.B. Hubbard and J. I Lunine}\affil{Department of Planetary Sciences, University of Arizona,
        Tucson, AZ 85721}
\author{M. Marley}\affil{Department of Astronomy, New Mexico State University,
                  Box 30001/Dept. 4500, Las Cruces NM 88003}
\author{T. Guillot}\affil{Department of Meteorology, University of Reading, P.O. Box 239,
Whiteknights,
Reading RG6 6AU,
United Kingdom}
\author{D. Saumon}\affil{Department of Physics and Astronomy, Vanderbilt University, Nashville, TN 37235}
\author{R.S. Freedman}\affil{Sterling Software,
                  NASA Ames Research Center, Moffett Field CA 94035}

\begin{abstract}

With the discovery of the companions of 51 Peg, 55 Cnc, $\tau$ Boo,
$\upsilon$ And, 70 Vir, 47 UMa, and Gl229, evolutionary and spectral models
of gas giants and/or brown dwarfs with masses from 0.3 through 60 times
that of Jupiter assume a new and central role in the emerging field of extrasolar
planetary studies.  In this contribution, we describe the structural, spectral, 
and evolutionary characteristics of such exotic objects, as determined by our recent
theoretical calculations.
These calculations can be used to establish direct search
strategies via SIRTF, ISO, and HST (NICMOS), and via various ground--based adaptive optics and
interferometric platforms planned for the near future.

\end{abstract}


\section{Introduction}

During the past year, scientists and the public at large have been
galvanized by the discovery of planets and brown dwarfs around nearby stars (Mayor \& Queloz 1995; Marcy \& Butler 1996; 
Oppenheimer \etal\ 1995; Butler \& Marcy 1996)
and by evidence
for ancient life on Mars (McKay \etal\ 1996).
These extraordinary findings have dramatically heightened
interest in the age--old questions of where we came from and whether we are unique in the cosmos.

Not unexpectedly, a general understanding of stellar and  planetary origins does not yet exist.
Planetary science has focussed over the past three decades
on the planets of our own solar system and on the chemical and physical
clues to their formation accessible by direct examination and sampling.
Astronomy has pushed
simultaneously outward to cosmology and inward to understand the formation
of stars and planets.
A long--term goal of theorists should be the integration
of these two realms and the creation of a new interdisciplinary science
of planets, stars, and life.

The new planet and brown dwarf discoveries (Table 1 and Figure 1) collectively have
an exceptional and unexpected range of masses, semi--major axes, eccentricities, and primaries.
The unpredicted variety among these new giant planets/brown dwarfs poses fundamental
questions about the origin of planetary and stellar systems.
While the new planets were all discovered by {\it indirect} means, it is only via {\it direct} detection
(imaging and spectroscopy)
that extrasolar planets can be adequately characterized and studied.
With this in mind, we have embarked upon a new and thorough theoretical study of the structure, spectra, colors, and evolution
of extrasolar giant planets and brown dwarfs (Marley \etal\ 1996; Burrows \etal\ 1997).

\begin{table}
\caption{The Bestiary of Brown Dwarfs and Extrasolar Giant Planets} \label{tbl-1}
\begin{center}\scriptsize
\begin{tabular}{crrrrrrrr}

Object&Star&M$_\star$ (M$_\odot$)&L$_\star$ (L$_\odot$)&d (pc)&M (M$_J$)&$a$ (AU)&P (days)&$e$ \\
\tableline

HD283750&K2V?&0.75?&0.2?&16.5&$\sgreat$50&$\sim$0.04&1.79&0.02 \\
$\tau$ Boo B&F7V&1.25&2.5&15&$\sgreat$3.87&0.046&3.313&0.018  \\
51 Peg B&G2.5V&1.0&1.0&15.4&$\sgreat$0.47&0.05&4.23&0.0 \\
HD98230&G0V&1.1&1.5&7.3&$\sgreat$37&0.05&3.98&0.0 \\
$\upsilon$ And B&F7V&1.25&2.5&17.6&$\sgreat$0.68&0.058&4.61&0.109 \\
55 Cnc B&G8V&0.85&0.5&13.4&$\sgreat$0.84&0.11&14.65&0.051 \\
HD112758&K0V&0.8&0.4&16.5&$\sgreat$35&$\sim$0.35&103.22&0.16 \\
HD114762&F9V&1.15&1.8&28&$\sgreat$10&0.38&84&0.25 \\
70 Vir B&G4V&0.95&0.8&18.1&$\sgreat$6.6&0.45&116.6&0.40 \\
HD140913&G0V&1.1&1.5&?&$\sgreat$46&$\sim$0.54&147.94&0.61 \\
HD89707&G1V&1.1&1.3&24.5&$\sgreat$54&0.69&198.25&0.95 \\
BD -04782&K5V&0.7&0.1&?&$\sgreat$21&$\sim$0.7&240.92&0.28 \\
HD110833&K3V&0.75&0.2&16&$\sgreat$17&$\sim$0.8&270.04&0.69 \\
HD217580&K4V&0.7&0.1&18.5&$\sgreat$60&$\sim$1.0&1.24 yrs&0.52 \\
HD18445&K2V&0.75&0.2&$\sim$20&$\sgreat$39&1.2&1.52 yrs&0.54 \\
16 Cyg B&G2.5V&1.0&1.0&22&$\sgreat$1.5&1.7&2.19 yrs&0.67 \\
47 UMa B&G0V&1.1&1.5&14.1&$\sgreat$2.4&2.1&2.98 yrs&0.03 \\
HD29587&G2V&1.0&1.0&42&$\sgreat$40&2.1&3.17 yrs&0.0 \\
Gl 411 B&M2V&0.4&0.02&2.52&$\sgreat$0.9&2.38&5.8 yrs&? \\
55 Cnc C&G8V&0.85&0.5&13.4&$\sgreat$5&3.8&$>$8 yrs&? \\
Jupiter&G2V&1.0&1.0&0.0&1.00&5.2&11.86 yrs&0.048 \\
Saturn&G2V&1.0&1.0&0.0&0.3&9.54&29.46 yrs&0.056 \\
Gl 229 B&M1V&0.45&0.03&5.7&30-55&$\sgreat$44.0&$\sgreat$400&? \\
\tableline

\end{tabular}
\end{center}

%\tablenotetext{a}{Sample footnote for Table~\ref{tbl-1}}

\end{table}

\begin{figure}
\vspace{3.50in}
\plotfiddle{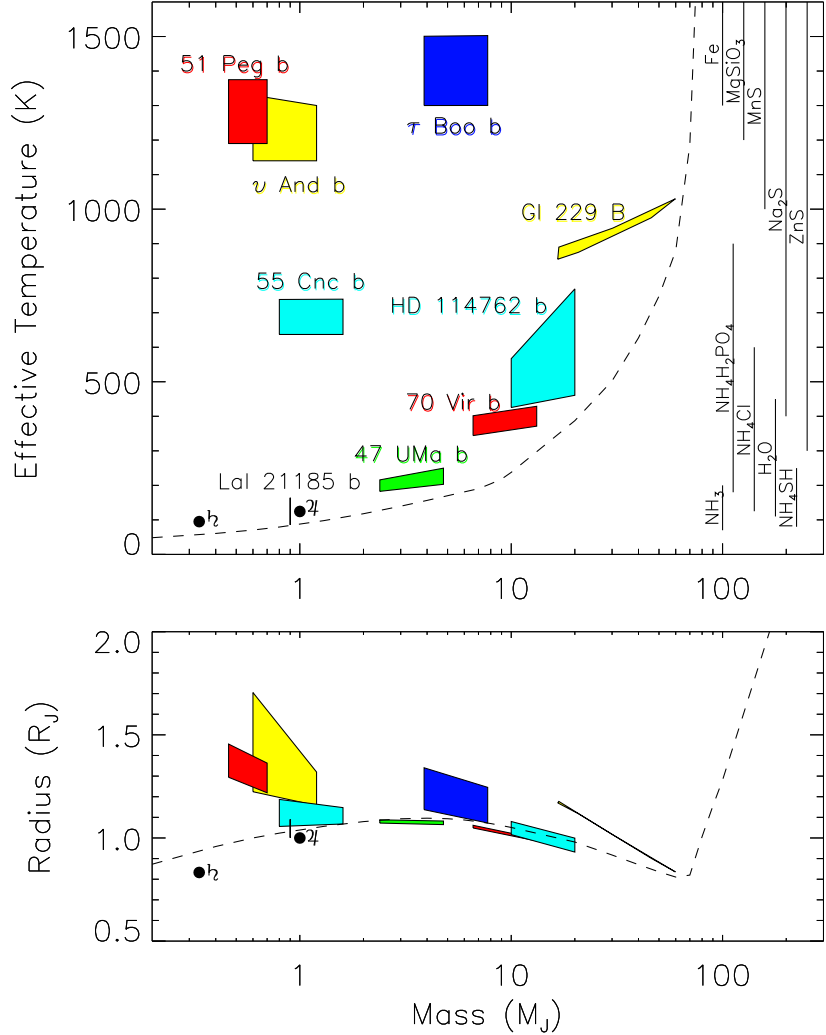}{3.5in}{0}{70}{70}{-200}{70}
\vspace{-2.0in}
\caption{Plots of the effective temperature (T$_{\rm eff}$) versus mass (in units
of \mj) (top panel) and of radius (in units of \rj) versus mass (bottom panel) 
for some of the newly--discovered planets and/or brown
dwarfs.
The effective temperature and radius ranges reflect the ambiguities in the Bond albedo, and the mass
ranges reflect a factor of two spread in sin(i) (except for Gl229 B).   The dotted curves are
10$^{10}$--year isochrones that nicely bound the family.  Stellar insolation effects have been included.
Lalande 21185 is depicted by a small vertical line in both panels.
To the right in the top panel the lines depict the range of T$_{\rm eff}$'s  for which the indicated species
condense out near the photosphere.
Hence, H$_2$O vapor features should be absent from the spectra of 70 Vir b
and 47 UMa b, since H$_2$O in those objects might be in the form of clouds.
In addition, Mg$_2$SiO$_4$ and Fe clouds might be in evidence in the spectra of $\tau$ Boo b, $\upsilon$ And b,
and 51 Peg b, if they are indeed gas giants.
}\label{fig-1}
\end{figure}

However, building upon our previous experience in the modeling of brown dwarfs and M stars,
we (Burrows \etal\ 1995; Saumon \etal\ 1996; and Guillot \etal\ 1996) first published grey models  
of extrasolar giant planets (EGPs\footnote[1]{We use this
shorthand for \undertext{E}xtrasolar \undertext{G}iant \undertext{P}lanet, but the terms ``exoplanet'' or ``super--jupiter''
are equally good.}).
This early generation of EGP models 
in the mass range 0.0003 \mo to 0.015 \mo ($\sim$ 0.3\mj to 15\mj)
was in aid of both NASA's and ESA's (Leger 1993) embryonic plans to search for extrasolar planets.
Figure 2 depicts the bolometric luminosity evolution of EGPs 
for ages from 10$^6$ to 5$\times10^{9}$ years.
Note that younger and more massive objects are {\it significantly} brighter and, hence, more easily detected.

%A sample figure appears in Figure~\ref{fig-2}.
\begin{figure}
\vspace{3.50in}
%\hbox to\hsize{\hfill\special{psfile=egpfig1.ps angle=-90 hscale=60 vscale=60 hoffset=-250 voffset=350}\kern+0in\hfill}
\hbox to\hsize{\hfill\includegraphics{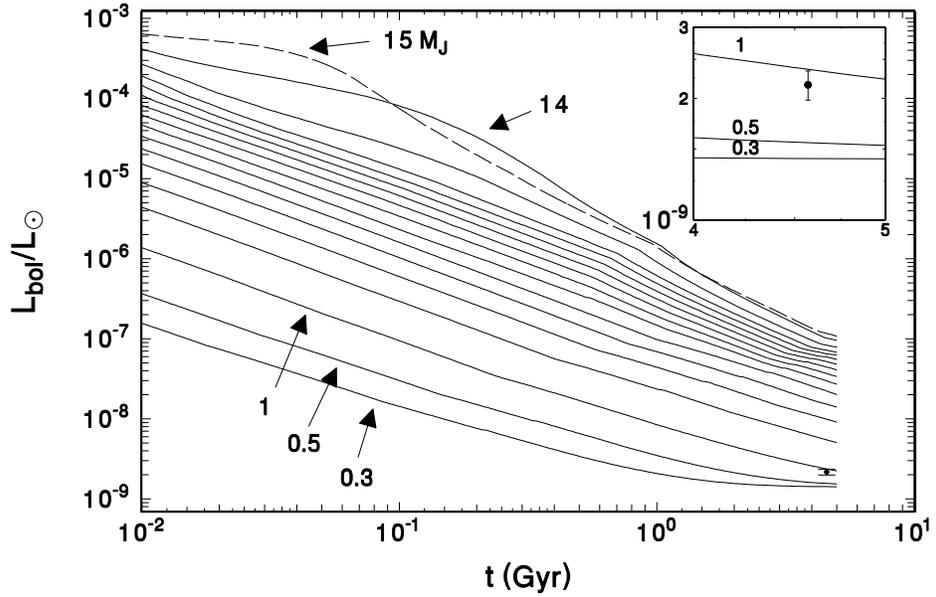}\kern+0in\hfill}
\vspace{.3in}
\caption{Bolometric luminosity
(\lbol) in solar units of a suite of EGPs placed at a
distance of 5.2 A.U. from a G2\Dwa star
versus  time ($t$) in Gyr.  The reflected luminosity is not included, but
the absorbed component is. At $t \sim 0.2\,$Gyr, the luminosity
of the 14\mj EGP exceeds that of the 15\mj EGP because of
late deuterium ignition.
The data point at $4.55\,$Gyr shows the
observed luminosity of Jupiter.  The 0.3\mj EGP
exhibits a strong effect of warming by the G2\Dwa primary star at
late stages in its evolution.
Although this model resembles
Saturn in mass, here it is placed at the distance of Jupiter from
its primary.
(The flattening in $L$ vs. $t$ for low masses and great ages is a
consequence of stellar insolation.)
The insert shows, on an expanded scale, the comparison
of our lowest-mass evolutionary trajectories with the present
Jupiter luminosity (from Burrows \etal\ 1995).}\label{fig-2}

\end{figure}

Some of the space platforms and new ground--based facilities that have or will
obtain relevant infrared and optical data include
the HST (WFPC2, NICMOS), the IRTF, the MMT 6.5-meter upgrade, the Large Binocular Telescope (LBT)
(planned for Mt. Graham), Keck's I and II (with HIRES), the European ISO, UKIRT,
and SIRTF, along with a large number of medium-- to large--size telescopes
optimized or employed in the near--infrared.
One project of Keck I and II,
under the aegis of NASA's ASEPS-0 (Astronomical Study of Extrasolar Planetary Systems)
program, will be to search for giant planets around nearby stars.
A major motivation for the Palomar Testbed Interferometer (PTI) supported by NASA
is the search for extrasolar planets.
Recently, Dan Goldin, the NASA administrator, outlined a program to detect planetary
systems around nearby stars that may become a future focus of NASA.
This vision is laid out in the {\it Exploration of Neighboring Planetary Systems (ExNPS)
Roadmap} (see also the ``TOPS'' Report, 1992).

\section{Early Calculations of the Evolution and Structure of Extrasolar Giant Planets}

Earlier work generally consisted of evolutionary
models of planets of 1$\, \mjj$ and below, beginning with
Graboske \etal\ (1975).
That work explored
the evolution of the low--mass objects Jupiter and Saturn from
an age of $10^7$ years to the present ($\sim$4.6 Gyr).  Working down
from higher masses, Grossman
and Graboske (1973) extended their calculations of
brown dwarf evolution to as low as
12 $\mjj$, but had to limit their study to ages less than about 0.1 Gyr.
Black (1980) used the results of Grossman \& Graboske (1973) and Graboske \etal\ (1975) to infer simple
power-law relations for the variation of luminosity $L$ and radius $R$ as
a function of mass $M_p$ and time $t$.  Black's relations are
roughly valid for objects close in mass to 1 $\mjj$ and close in
age to 4.5 Gyr.  However,
Black's formulas become inaccurate at
earlier ages and at larger masses.

EGPs will radiate in the optical by reflection and in the
infrared by the thermal emission of both absorbed stellar light and the planet's
own internal energy.
In Burrows \etal\ (1995), Saumon \etal\ (1996), and Guillot \etal\ (1996), we 
evolved EGPs with masses ($M_p$) from 0.3\mj $\,$(the mass of Saturn)
through 15\mj and included 
the effects of ``insolation'' by a central star. 
Giant planets may form preferentially near 5 A.U. (Boss 1995), but as the new data dramatically affirm,
a broad range of $a$\,s can not be excluded.   
Whether a 15\mj object is a planet or a brown dwarf is largely a semantic issue, though one might
distinguish gas giants and brown dwarfs by their
mode of formation (e.g. in a disk or ``directly'').  Physically, compact hydrogen-rich objects with masses from 0.00025 \mo$\,$ through
0.25 \mo$\,$ form a continuum.  However, EGPs
above $\sim$13\mj$\,$ do burn deuterium for up to 10$^{8}$ years.

\section{New Evolutionary and Spectral Models of EGPs}

On the same October day on which 51 Peg B was reported (Mayor \& Queloz 1995), 
a Caltech team announced the {\it direct} detection of the brown dwarf
Gliese 229 B (Geballe \etal\ 1995; Oppenheimer \etal\ 1995; Nakajima \etal\ 1995; Matthews \etal\ 1996).
Using a prototype adaptive optics system on the old MMT, Roger Angel of Steward Observatory
obtained a K--band image of Gl 229 B, thus confirming its detection.
This object has an estimated luminosity of $6.4 \pm 0.6 \times 10^{-6}
L_{\odot}$, an effective temperature
below 1200 K, and clear signatures of methane and H$_2$O vapor in its spectrum.
Since there can be no stars cooler than 1700 K, with luminosities below
$5 \times 10^{-5} L_{\odot}$, or
with methane bands, Gl 229 B's status as one of the long--sought brown dwarfs is now unimpeachable.

To constrain the properties of the brown dwarf Gl229 B,
we (Marley \etal\ 1996)
constructed a grid of non--grey model
atmospheres with $T_{\rm eff}$ ranging from 700 to 1200 K and
$10^{4}<g<3\times10^{5}$ cm s$^{-2}$.
We assumed a standard solar composition for the bulk of
the atmosphere.
Refractory elements (for example Fe, Ti, and silicates) condense deep in the
atmosphere for $T_{\rm eff} \approx 1000$ K, and thus
have negligible gas--phase abundance near the photosphere, as
is also true in the atmosphere of Jupiter.
For an atmosphere similar to that of Gl229 B, chemical equilibrium
calculations indicate that C, N, O, S, and P
are found mainly in the form of methane (CH$_4$), ammonia (NH$_3$),
water (H$_2$O), hydrogen sulfide (H$_2$S), and phosphine (PH$_3$),
respectively.   

Our model atmospheres incorporate
opacities due to collision--induced absorption by H$_2$-H$_2$
(Borysow \& Frommhold 1990)
and H$_2$-He (Zheng \& Borysow 1995, and references therein),
free-free absorption by H$_2^{-}$
(Bell 1980),
bound-free absorption by H$^{-}$
(John 1988),
and Rayleigh scattering.
The absorptions of NH$_3$, CH$_4$, and PH$_3$ are calculated using the HITRAN
data base (Hilico \etal\ 1992)
with corrections and extensions.
Additional tabulations
are used where necessary for $\rm CH_4$ (Strong \etal\ 1993), especially shortwards of $1.6\,\rm \mu m$.  Data for
H$_2$O and H$_2$S are computed from a direct numerical diagonalization
(Wattson \& Rothman 1992; Schwenke \etal\ 1997).
Absorption by CO (Pollack {\it et al.} 1993)
and PH$_3$ opacity is included in the
spectral models, but not in the temperature profile computation.
The baseline models assume that the atmosphere is free of clouds.
This assumption must be reconsidered in the future (see Figure 1).

For the temperature profile computation, molecular opacity is treated using
the k--coefficient method
(Goody \etal\ 1989).
After a radiative--equilibrium temperature profile is found, the
atmosphere is iteratively adjusted to self-consistently solve
for the size of the convection zones, given the specified internal
heat flux.  With the radiative--convective temperature--pressure profiles,
high-resolution synthetic spectra can be generated by solving the
radiative transfer equation.

Figure 3 shows a comparison of a theoretical spectrum of Gl229 B with the UKIRT spectrum from
1 to 2.5 $\mu$ms.  Including stray light in the optics improves the fit
in the H$_2$O troughs dramatically and a T$_{\rm eff}$ near 1000 K is indicated.  This
is far below the stellar edge T$_{\rm eff}$ of $\sim$1750 K for solar metallicity,
indicating that Gl229 B is a true brown dwarf with a mass near 30$\pm$10\mj.

\begin{figure}
\vspace{3.5in}
%\hbox to\hsize{\hfill\special{psfile=gl229.scat.ps angle=-90 hscale=60 vscale=60 hoffset=-230 voffset=325}\kern+0in\hfill}
\hbox to\hsize{\hfill\includegraphics{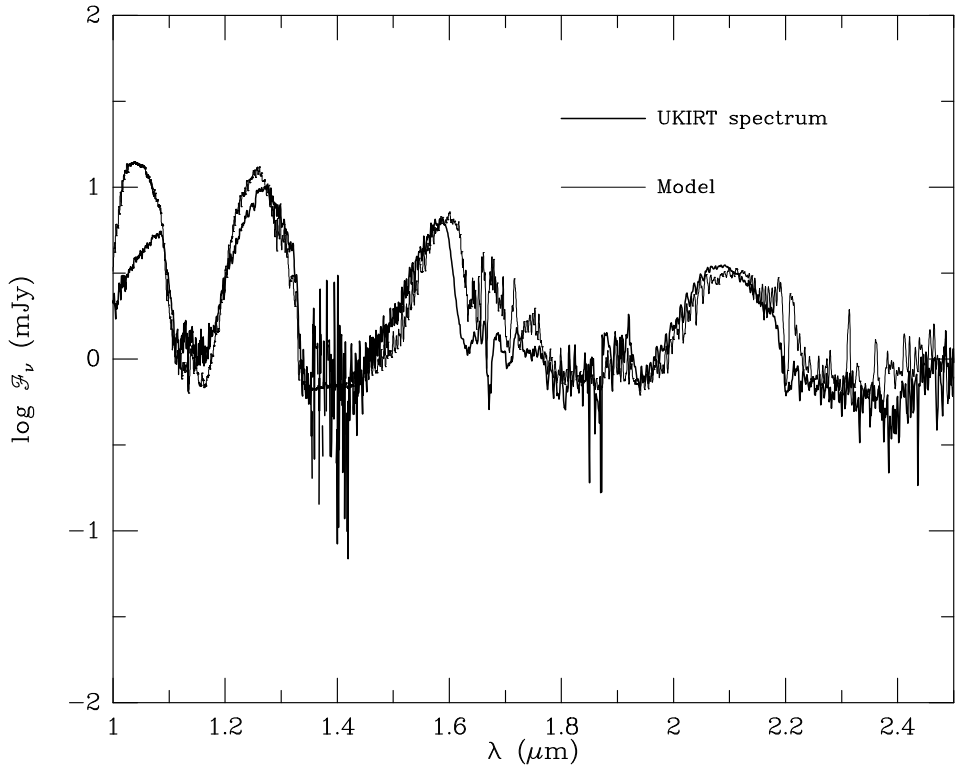}\kern+0in\hfill}
\vspace{.5in}
\caption{A comparison of the observed UKIRT spectrum of Gl229 B with a recent theoretical spectrum,
including the effects of stray light in the optics.  The scattered light fills in the troughs
that theory says would otherwise be very much deeper.  The discrepancy at the shorter wavelengths
may be a consequence of Rayleigh scattering by ``clouds'' in the atmosphere of Gl229 B and is not 
due to the presence of TiO or metal hydride absorbers present in M dwarf atmospheres. 
}\label{fig-3}
\end{figure}

\begin{figure}
\vspace{3.50in}
\plotfiddle{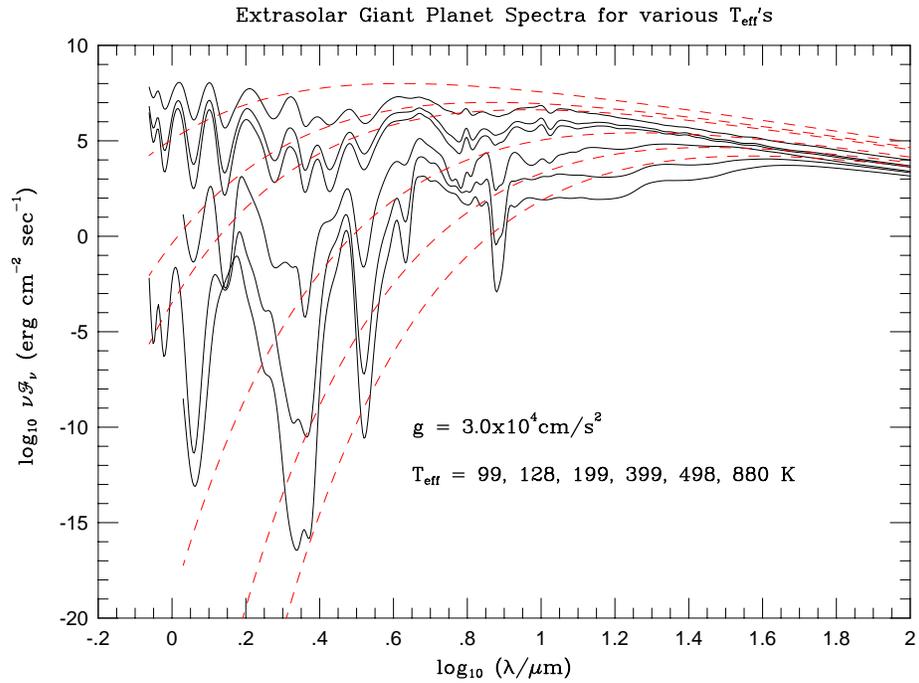}{3.5in}{90}{50}{50}{190}{200}
\vspace{-3.0in}
\caption{A comparison with the corresponding blackbody spectra of some preliminary
theoretical spectra from 99 to 880 Kelvin for EGPs and brown dwarfs at $3\times10^4$ cm s$^{-2}$, {\it in isolation}.
Plotted is $\nu\cal{F}_{\nu}$ versus $\lambda$ in microns, where $\cal{F}_{\nu}$ is the flux
from the object's surface, not the flux in a detector.  The latter is easily derived knowing the radius
as a function of T$_{\rm eff}$ and gravity.
Sufficient flux is redistributed from the longer wavelengths to the near infrared
to force a dramatic reappraisal of search strategies.  The J, H and M bands are significantly
enhanced over the mid--infrared and the K band, though the K band is near the blackbody value
for T$_{\rm eff}$ below $\sim$500 K
(from Burrows \etal\ 1997).
}\label{fig-4}
\end{figure}

As part of our recent theoretical explorations after the Gl229 B campaign,  
we have found significant flux enhancements relative to blackbody values in
the window at 4--5 $\mu$m and in the J (1.2 $\mu$m) and H(1.6 $\mu$m) bands for T$_{\rm eff}$s  from 100 
through 1300 K.  At temperatures below $\sim$600 K, 
the widely-used K band at 2.2 $\mu$m is greatly suppressed {\it relative to the J and H bands} by strong CH$_4$ and
H$_2-$H$_2$ absorption features.  The  J, H, and M band enhancements are a consequence of the opacity 
holes in those bands and the significant redistribution
of flux caused predominantly by pressure--induced H$_2$ absorption longward of 5 $\mu$ms and strong absorption
features of CH$_4$ and H$_2$O in the near and mid--infrared.
Figure 4 depicts some preliminary theoretical spectra for objects from 99 to 880 K at a gravity of $3\times10^{4}$ cm s$^{-2}$
(Burrows \etal\ 1997).
Infrared colors derived from these spectra are easily distinguished from those of M dwarfs.
In particular, J-K and H-K  colors get {\it bluer} with decreasing T$_{\rm eff}$ and are shifted relative
to those for M dwarfs and hot brown dwarfs by many magnitudes. 
The gravity dependence of the emergent spectra is weak.
For comparison, in Figure 4 the corresponding blackbody spectra are provided.  
While the giant planets
of the solar system can be helpful guides at the low-$T_{\rm eff}$ limit of our
calculations, the range demonstrated by their spectra warns us
against simple generalizations. The extremely non--blackbody nature of both
brown dwarfs and extrasolar giant planets is manifest and rather startling.   Search strategies should be redesigned
accordingly.

\section{The Future}

The present pace of giant planet discovery and NASA's future plans
for planet searches suggest that
many more objects in the Jovian mass range (and above)
will be identified and subject to spectroscopic examination. 
Spectra and colors are worth little if they can not be
attached to masses, ages, and compositions.
We have already preformed the first rudimentary calculations
of the atmospheres, evolution, and spectra of extrasolar giant planets, in a variety of
stellar environments.  
As our recent work on Gl229 B and Figure 4 show, the blackbody assumption can be
orders--of--magnitude off in the J, H, L$^{\prime}$, and M bands.  
Therefore, a complete and self--consistent theory of the evolution, colors, spectra, and structure of extrasolar giant planets
is a crucial prerequisite
for any credible direct search for planets around nearby stars.
The new science of extrasolar planets, a merger of both planetary and stellar astronomy,
is rapidly being born.

\acknowledgments

The authors would like to thank Jim Liebert, Roger Angel, Christopher Sharp, 
Andy Nelson, David Sudarsky, Willy Benz, George Gatewood, Geoff Marcy, and David Trilling
for many simulating and fruitful conversations 
during the last hectic 18 months.  They would also like to acknowledge the
support of NASA via grant NAGW--2817 and of the NSF via grant
AST93-18970.

%
% Observe the "standard" order for bibliographic material: author name(s),
% publication year, journal name, volume, and page number for articles.
%


\begin{references}

\reference Bell, K.L. 1980, {\it J.  Phys. B}, 13 1859
\reference Borysow, A. \& Frommhold, L. 1990, \apjl, 348, L41
\reference Black, D.C. 1980, Icarus, 43, 293
\reference Boss, A.P. 1995, Science, 267, 360
\reference Burrows, A., Hubbard, W.B., Saumon, D., \& Lunine, J.I. 1993, \apj, 406, 158
\reference Burrows, A., Saumon, D., Guillot, T., Hubbard, W.B., \& Lunine, J.I. 1995, 
                Nature, 375, 299
\reference Burrows, A., Hubbard, W. B., Marley, M., Guillot, T., Lunine, J. I., Saumon, D., \&  Freedman, R. S. 1997, in preparation
\reference Butler, R. P. \& Marcy, G. W. 1996, \apjl, 464, L153
\reference ExNPS: A Road Map for the Exploration of Neighboring Planetary Systems, JPL Publication 96--22,
August 1996
\reference Geballe, T. R., Kulkarni, S. R., Woodward, C. E., \& Sloan, G. C. 1996, \apjl, 467, L101
\reference Goody, R., West, R., Chen, L., \& Crisp, D. 1989,
{\sl J. Quant. Spectr. Radiat. Transfer}, 42, 539
\reference Graboske, H.C., Jr., Pollack, J.B., Grossman, A.S. \&  Olness, R.J. 1975, \apj, 
                   199, 265 
\reference Grossman, A.S. \& Graboske, H.C., Jr. 1973, \apj, 180, 195  
\reference Guillot, T., Burrows, A., Hubbard, W.B.,  Lunine, J.I., \&  Saumon, D. 1996, \apjl, 459, L35
\reference Hilico, J.C., Loete, M., \& Brown, L.R., Jr. 1992,  {\it J. of Mol.
Spectr.}, 152, 229
\reference ISO Observer's Manual Version 2.0, 31 March 1994, prepared by the ISO Science
               Operations Team, p. 6
\reference John, T.L. 1988, \aa, 193, 189
\reference Leger, A., \etal\ 1993, Darwin Mission Concept, proposal to ESA 
\reference Marcy, G. W. \& Butler, R. P. 1996, \apjl, 464, L147
\reference Marley, M. S., Saumon, D.,  Guillot, T., Freedman, R., Hubbard, W. B., Burrows, A., \& Lunine, J. I. 1996, 
Science, 272, 1919
\reference Matthews, K., Nakajima, T., Kulkarni, S. R., \& Oppenheimer, B. R. 1996, submitted to \apj
\reference Mayor, M. \& Queloz, D. 1995, Nature, 378, 355
\reference McKay, D. S., \etal\ 1996, Science, 273, 924
\reference Nakajima, T., {\it et al.} 1995, Nature, 378, 463
\reference Oppenheimer, B. R., Kulkarni, S. R., Matthews, K., \& Nakajima, T. 1995, Science, 270, 1478
\reference Pollack, J.,  {\it et al.} 1993,
{\it Icarus}, 103, 1
\reference Saumon, D., Hubbard, W. B., Burrows, A., Guillot, T., Lunine, J. I., \& Chabrier, G. 1996, \apj, 460, 993
\reference Strong, K., Taylor, F.W., Calcutt, S.B., Remedios, J.J.,
\& Ballard, J. 1993, {\it J. Quant. Spectr. Radiat. Transfer}, 50, 363
\reference Schwenke, R. \etal\ 1997, in preparation
\reference TOPS: Toward Other Planetary Systems, NASA Solar System Exploration
Division, Washington, D.C., 1992
\reference Wattson, R.B., and Rothman L.S. 1992, {\sl J. Quant. Spectr. Radiat. Transfer}, 48, 763
\reference Zheng, C. \& Borysow, A. 1995, Icarus, 113, 84

\end{references}
\end{document}